\documentclass{emulateapj}

\usepackage{color}
\usepackage{multirow}

\usepackage{natbib}
\shorttitle{ABSENCE OF EVIDENCE IS NOT EVIDENCE OF ABSENCE}
\shortauthors{Cooper et al.}

\begin{document}


%\title{Overstating an Apparent Lack of Environmental
%  Dependence: A Cautionary Tale from zCOSMOS, VVDS, and DEEP2}
\title{Absence of evidence is not evidence of absence: the
  color-density relation at fixed stellar mass persists to
  $\lowercase{z} \sim 1$\footnotemark[\dag]}


\author{
Michael C.\ Cooper\altaffilmark{1,2},
Alison L.\ Coil\altaffilmark{3,4},
Brian F.\ Gerke\altaffilmark{5},
Jeffrey A.\ Newman\altaffilmark{6},
Kevin Bundy\altaffilmark{7,8},
Christopher J.\ Conselice\altaffilmark{9},
Darren J. Croton\altaffilmark{10},
Marc Davis\altaffilmark{7,11},
S.\ M. Faber\altaffilmark{12},
Puragra Guhathakurta\altaffilmark{11},
David C.\ Koo\altaffilmark{12},
Lihwai Lin\altaffilmark{13},
Benjamin J.\ Weiner\altaffilmark{1},
Christopher N.\ A.\ Willmer\altaffilmark{1},
Renbin Yan\altaffilmark{14}
}

\footnotetext[\dag]{Some of the data presented herein
  were obtained at the W.M.\ Keck Observatory, which is operated as a
  scientific partnership among the California Institute of Technology,
  the University of California and the National Aeronautics and Space
  Administration. The Observatory was made possible by the generous
  financial support of the W.M.\ Keck Foundation.}

\altaffiltext{1}{Steward Observatory, University of Arizona;
  cooper@as.arizona.edu, cnaw@as.arizona.edu, bjw@as.arizona.edu}

\altaffiltext{2}{Spitzer Fellow}

\altaffiltext{3}{Center for Astrophysics and Space Sciences,
  University of California, San Diego}

\altaffiltext{4}{Alfred P.\ Sloan Foundation Fellow}

\altaffiltext{5}{Kavli Institute for Particle Astrophysics and
  Cosmology, Stanford Linear Accelerator Center;
  bgerke@slac.stanford.edu}

\altaffiltext{6}{Department of Physics and Astronomy, University of
  Pittsburgh; janewman@pitt.edu}

\altaffiltext{7}{Department of Astronomy, University of California,
  Berkeley; kbundy@astro.berkeley.edu, marc@astro.berkeley.edu}

\altaffiltext{8}{Hubble Fellow}

\altaffiltext{9}{School of Physics and Astronomy, University of
  Nottingham; conselice@nottingham.ac.uk}

\altaffiltext{10}{Center for Astrophysics and Super Computing,
  Swinburne University of Technology; dcroton@astro.swin.edu.au}

\altaffiltext{11}{Department of Physics, University of California, Berkeley}

\altaffiltext{12}{UCO/Lick Observatory and Department of Astronomy and
  Astrophysics, University of California, Santa Cruz;
  faber@ucolick.org, koo@ucolick.org, raja@ucolick.org}

\altaffiltext{13}{Institute of Astronomy and Astrophysics, Academia
  Sinica; lihwailin@asiaa.sinica.edu.tw}

\altaffiltext{14}{Department of Astronomy \& Astrophysics, University
  of Toronto; yan@astro.utoronto.ca}

\begin{abstract}
  We use data drawn from the DEEP2 Galaxy Redshift Survey to
  investigate the relationship between local galaxy density, stellar
  mass, and rest-frame galaxy color. At $z \sim 0.9$, we find that the
  shape of the stellar mass function at the high-mass $(\log_{10}({\rm
    M}_{*} / h^{-2}\ {\rm M}_{\sun}) > 10.1)$ end depends on the local
  environment, with high-density regions favoring more massive
  systems. Accounting for this stellar mass-environment relation
  (i.e., working at fixed stellar mass), we find a significant
  color-density relation for galaxies with $10.6 < \log_{10}({\rm
    M}_{*} / h^{-2}\ {\rm M}_{\sun}) < 11.1$ and $0.75 < z <
  0.95$. This result is shown to be robust to variations in the sample
  selection and to extend to even lower masses (down to
  $\log_{10}({\rm M}_{*} / h^{-2}\ {\rm M}_{\sun}) \sim 10.4$). We
  conclude by discussing our results in comparison to recent works in
  the literature, which report no significant correlation between
  galaxy properties and environment at fixed stellar mass for the same
  redshift and stellar mass domain. The non-detection of environmental
  dependence found in other data sets is largely attributable to their
  smaller samples size and lower sampling density, as well as
  systematic effects such as inaccurate redshifts and biased analysis
  techniques. Ultimately, our results based on DEEP2 data illustrate
  that the evolutionary state of a galaxy at $z \sim 1$ is not
  exclusively determined by the stellar mass of the galaxy. Instead,
  we show that local environment appears to play a distinct role in
  the transformation of galaxy properties at $z > 1$.
 
\end{abstract}


\section{Introduction}
\label{sec_intro}

With the emergence of large spectroscopic surveys at intermediate
redshift, such as the DEEP2 Galaxy Redshift Survey \citep{davis03,
  newman09}, the VIMOS VLT Deep Survey \citep[VVDS,][]{lefevre05}, and
zCOSMOS \citep{lilly07}, robust systematic studies of galaxy
clustering are now able to be extended out to $z \sim 1$
\citep[e.g.,][]{coil04a, coil06, meneux08, meneux09}. In particular,
these unprecedented data sets have enabled the local density of
galaxies (i.e., on scales of $\sim \! 1$--$2$ $h^{-1}$ Mpc) to be
statistically measured over a broad and continuous range of
environments extending from voids to rich groups and poor clusters at
$z \sim 1$ \citep[e.g.,][]{cooper05, cooper06, cucciati06,
  kovac10}. Moreover, individual galaxy groups at $z \sim 1$ are now
able to be reliably identified in redshift space, enabling the
``field'' population to be studied in relation to the ``group'' or
``cluster'' population \citep[e.g.,][]{gerke05, knobel09, cucciati09}.

Capitalizing on the ability to characterize environment over half of
the age of the Universe, many studies have addressed the role of
environment in galaxy evolution at $z < 1$ by studying the
relationship between environment and galaxy properties such as
rest-frame color and morphology \citep[e.g.,][]{gerke07, elbaz07,
  capak07, cooper08}. For example, using data drawn from the DEEP2
survey, \citet{cooper06} showed that all features of the global
correlation between galaxy color and environment measured locally are
already in place at $z \sim 1$ \citep[see also][]{coil08}. This work
emphasized that physical processes specific to clusters (such as
ram-pressure stripping and harassment) are not required to explain the
color-density relation at $z \sim 1$, given the lack of such extreme
environments in the DEEP2 sample.

However, as highlighted by several recent analyses, nearly all of the
early studies addressing the relationships between environment and
galaxy properties at intermediate redshift primarily utilized
luminosity-selected samples and not stellar mass-selected samples. A
number of recent papers have concluded that while there is evidence
for a color-density relation at fixed luminosity at $z \sim 1$, they
find no color-density relation at fixed mass. For example,
\citet{scodeggio09} employ data from VVDS to examine the relationship
between rest-frame $B-I$ color and environment as a function of
redshift within fixed bins of stellar mass. They find no evidence for
a color-density relation at fixed mass over the entire redshift $(0.2
< z < 1.4)$ and stellar mass $(9 < \log_{10}({\rm M}_{*}) < 11)$
ranges probed by their data. Using zCOSMOS to characterize the local
environment, \citet{tasca09} also find that at $0.5 < z < 1$ there is
no variation in galaxy morphology (i.e., early- versus late-type) as a
function of local galaxy density at stellar masses of $\log_{10}({\rm
  M}_{*} / h^{-2}\ {\rm M}_{\sun}) > 10.5$. From these results, both
of the aforementioned studies conclude that the properties (i.e.,
color and morphology) of massive galaxies are independent of
environment at $z \sim 1$.

Given the substantial noise in all current measures of environment
(both statistical measures of the local galaxy density and group/field
catalogs), intrinsic correlations between galaxy properties and
environment can easily be smeared out, such that no significant trends
with environment are detectable in the data. That is, it is far easier
to smear out a correlation with environment than it is to erroneously
detect one as, generally, measures of local density are dependent on
redshift measurements of neighboring galaxies, while a given galaxy's
properties are measured from its own imaging and spectroscopic data.
It is difficult to imagine a systematic effect that would yield a
false correlation between independently determined quantities.

Conclusions based on the lack of an apparent environment dependence
must therefore be drawn with caution. In this paper, we use data from
the DEEP2 survey to investigate the significance or robustness of the
recent, aforementioned conclusions inferred from studies of galaxy
environments at intermediate redshift. In Section \ref{sec_data}, we
describe our data set, with results and discussion presented in
Sections \ref{sec_results} and \ref{sec_disc}, respectively.
Throughout, we employ a $\Lambda$CMD cosmology with $w = -1$,
$\Omega_m = 0.3$, $\Omega_{\Lambda} = 0.7$, and a Hubble parameter of
$H_0 = 100\ h$ km s$^{-1}$ Mpc$^{-1}$. All magnitudes are on the AB
system.

\section{Data}
\label{sec_data}

Among the current generation of deep spectroscopic redshift surveys at
$z \sim 1$, the DEEP2 Galaxy Redshift Survey provides the largest
sample of accurate spectroscopic redshifts, the highest-precision
velocity information, and the highest sampling density.\footnote{Note
  that the sampling density for a survey is defined to be the number
  of galaxies with an accurate redshift measurement per unit of
  comoving volume and \emph{not} the number of galaxies targeted down
  to an arbitrary magnitude limit.}  Altogether, these attributes make
the DEEP2 survey the best existing spectroscopic data set with which
to characterize local environment. In this paper, we utilize a sample
of $23,037$ galaxies with accurate redshifts \citep[quality $Q = 3$ or
4 as defined by][]{newman09} in the range $0.75 < z < 1.25$ and drawn
from all four of the DEEP2 survey fields.

For each galaxy in the DEEP2 sample, rest-frame $U-B$ colors and
absolute $B$-band magnitudes, $M_B$, are calculated from CFHT $BRI$
photometry \citep{coil04b} using the $K$-correction procedure
described in \citet{willmer06}. For a portion of the DEEP2 catalog,
stellar masses may be calculated by fitting spectral energy
distributions (SEDs) to WIRC/Palomar $J$- and $K_s$-band photometry in
conjunction with the DEEP2 $BRI$ data, according to the prescriptions
described by \cite{bundy05, bundy06}. However, the near-infrared
photometry, collected as part of the Palomar Observatory Wide-field
Infrared \citep[POWIR,][]{conselice08} survey, does not cover the
entire DEEP2 survey area, and often faint blue galaxies at the
high-$z$ end of the DEEP2 redshift range are not detected in
$K_s$. Because of these two effects, the stellar masses of
\citet{bundy06} have been used to calibrate stellar mass estimates for
the full DEEP2 sample that are based on rest-frame $M_B$ and $B-V$
values derived from the DEEP2 data in conjunction with the expressions
of \citet{bell03}. We empirically correct these stellar mass estimates
to the \citet{bundy06} measurements by accounting for a mild color and
redshift dependence \citep{lin07}; where they overlap, the two stellar
masses have an rms difference of approximately $0.3$ dex after this
calibration.

To characterize the local environment, we compute the projected
third-nearest-neighbor surface density $(\Sigma_3)$ about each galaxy
in the DEEP2 sample, where the surface density depends on the
projected distance to the third-nearest neighbor, $D_{p,3}$, as
$\Sigma_3 = 3 / (\pi D_{p,3}^2)$. In computing $\Sigma_3$, a velocity
window of $\pm 1000$ km s$^{-1}$ is utilized to exclude foreground and
background galaxies along the line of sight. Varying the width of this
velocity window (e.g., using $\pm 1500$ km s$^{-1}$) or tracing
environment according to the projected distance to the fifth-nearest
neighbor has no significant effect on our results. In the tests of
\citet{cooper05}, this projected $n^{\rm th}$-nearest-neighbor
environment estimator proved to be the most robust indicator of local
galaxy density for the DEEP2 survey.

To correct for the redshift dependence of the DEEP2 sampling rate,
each surface density value is divided by the median $\Sigma_3$ of
galaxies at that redshift within a window of $\Delta z = 0.04$;
correcting the measured surface densities in this manner converts the
$\Sigma_3$ values into measures of overdensity relative to the median
density (given by the notation $1 + \delta_3$ here) and effectively
accounts for the redshift variations in the selection rate
\citep{cooper05}. Finally, to minimize the effects of edges and holes
in the survey geometry, we exclude all galaxies within $1$ $h^{-1}$
comoving Mpc of a survey boundary, reducing our sample to $17,767$
galaxies in the redshift range $0.75 < z < 1.25$.

\begin{figure}[h!]
\centering
\plotone{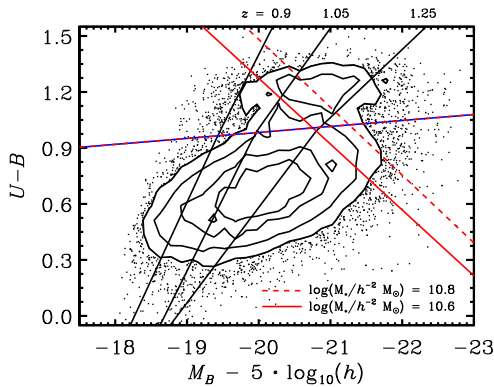}
\caption{the rest-frame $U-B$ versus $M_B$ color-magnitude
  distribution for DEEP2 galaxies in the spectroscopic sample within
  the redshift range $0.75 < z < 1.25$. The three solid black vertical
  lines show the completeness limit of the survey at $z = 0.9$,
  $1.05$, and $1.25$, while the solid and dashed red lines show lines
  of constant stellar mass corresponding to $\log_{10}({\rm
    M}_{*}/h^{-2}\ {\rm M}_{\sun}) = 10.6$ and $10.8$, respectively.
  The dashed blue/red horizontal line shows the division between the
  red sequence and the blue cloud as given by Equation 19 from
  \citet{willmer06}.}
\label{cmd_fig}
\end{figure}

To investigate the relationship between galaxy properties and
environment within the high-mass segment of the galaxy population, we
define a subsample of galaxies with stellar mass in the range $10.6 <
\log_{10}({\rm M}_{*} / h^{-2}\ {\rm M}_{\sun}) < 11.1$ and a redshift
of $0.75 < z < 1.05$. The median redshift for this subsample of $1586$
galaxies is $0.89$ and the median stellar mass is $\log_{10}({\rm
  M}_{*} / h^{-2}\ {\rm M}_{\sun}) \sim 10.8$. We restrict the
subsample in redshift to a range over which the DEEP2 redshift
selection function is relatively flat. However, over this redshift
range the sample is incomplete at the adopted mass limit. For example,
at $z=0.9$ the $R_{\rm AB}=24.1$ magnitude limit of DEEP2 includes all
galaxies with stellar mass $> \!  10^{10.8}\ {\rm M}_{*}/h^{-2}\ {\rm
  M}_{\sun}$ independent of color, but preferentially misses red
galaxies at lower masses (cf.\ Figure \ref{cmd_fig}).

Figure \ref{cmd_fig} shows the distribution of galaxies in
color-magnitude space for the entire DEEP2 sample at $0.75 < z <
1.25$, with lines of constant stellar mass overlaid. As shown in many
previous studies \citep[e.g.,][]{bell01, hogg03, cooper08}, selecting
by stellar mass as opposed to optical luminosity selects a different
portion of the galaxy population due to the dependence of stellar
mass-to-light ratio on rest-frame color. A $B$-band
luminosity-selected sample is biased towards lower stellar masses for
blue galaxies and higher stellar masses for red galaxies.

Now, both locally and at $z \sim 1$, the local environment of galaxies
on the red sequence has been shown to depend on luminosity such that
more luminous systems favor higher-density regions on average
\citep{blanton05, cooper06}. Thus, relative to a random population of
blue galaxies, red galaxies of equivalent stellar mass will be (less
luminous and) in lower-density environments than red galaxies selected
to have the same luminosity. That is, taking into account the known
correlation between local environment and galaxy luminosity, for a
galaxy sample selected to have fixed stellar mass (versus fixed
luminosity) one expects to find a weaker color-density
relation. However, it remains unclear whether or not the correlation
between galaxy environment and rest-frame color found at fixed
luminosity at $z \sim 1$ \citep{cooper06} is entirely a projection of
the correlation between environment and luminosity in concert with the
dependence of stellar mass-to-light ratio on color such that at fixed
stellar mass there is no color-density relation. As discussed in \S
\ref{sec_intro}, several recent analyses have arrived at this
conclusion.

\section{Analysis}
\label{sec_results}

As highlighted in \S \ref{sec_intro}, the goal of this work is to
study the relationship between galaxy color and local environment at
fixed stellar mass. Recent analyses using a variety of data sets have
approached this task using galaxy samples in stellar-mass-selected
bins across a range of environments. Here, we first investigate
whether this methodology is appropriate. Finding it is not --- that
is, finding that the stellar mass function is different in different
environments --- we then apply improved techniques which account for
the correlation between stellar mass and environment in an examination
of the color-density relation at $z \sim 1$. In contrast to recent
results in the literature, we find that there is a significant
color-density relation at fixed stellar mass. The most likely reason
for this discrepancy is that both larger measurement errors and
systematics in other data sets smear out the underlying environment
dependence. Finally, we emphasize that absence of evidence is not
evidence of absence;\footnote{We note that this turn of phrase, which
  we utilize in the title to this work, is a commonplace made popular
  by Carl Sagan among others in discussing a particular logical
  fallacy related to the misinterpretation of non-significant
  scientific results.} that is, not finding a correlation between
color and environment at fixed stellar mass does not directly indicate
an intrinsic lack of correlation. By making such an assumption, the
conclusions of several recent analyses have been rendered difficult to
interpret.

\subsection{The Relationship between Stellar Mass and Environment}

In order to study the relationship between galaxy properties and
environment at fixed stellar mass, the galaxy sample under study is
often restricted to a narrow range in stellar mass such that
correlations between stellar mass and environment are
negligible. However, at intermediate redshift, sample sizes are
generally limited in number such that using a particularly narrow
stellar mass range (e.g., $\sim \!  0.1$--$0.2$ dex in width)
significantly reduces the statistical power of the sample. For this
reason, broader stellar mass bins (e.g., $\sim \! 0.5$ dex) are
commonly employed \citep[e.g.,][]{scodeggio09, iovino09, patel09,
  kovac09, maltby10}. Now, if the shape of the stellar mass function
depends on environment, then the typical stellar mass within a broad
mass bin may differ significantly from one density regime to
another. Such an effect would clearly impact the ability to study the
color-density relation at fixed stellar mass.

To investigate the level to which we are sensitive to a stellar
mass-environment relation within our adopted stellar mass bin, we
select those galaxies within the top $10\%$ of the overdensity
distribution for all galaxies at $10.6 < \log_{10}({\rm M}_{*} /
h^{-2}\ {\rm M}_{\sun}) < 11.1$ and $0.75 < z < 1.05$ --- a subsample
of $159$ galaxies. From the corresponding bottom $50\%$ of the
overdensity distribution, we draw $1000$ random galaxy subsamples
(each consisting of $159$ galaxies), where each subsample is selected
to match the redshift distribution of the galaxies in the high-density
subsample. 

By matching in redshift, we remove the projection of any possible
residual correlation between our environment measurements and redshift
(in concert with the known redshift dependence of the survey's
stellar-mass limit) onto the observed stellar mass-environment
relation. This effect appears to be impacting the analysis of
\citet{iovino09}. As shown in Figure 12 of that work, the typical
redshift of ``isolated'' galaxies (i.e., galaxies in low-density
environments) is systematically skewed towards higher $z$ relative to
the sample of ``group'' members (i.e., galaxies in high-density
environments). This effect is likely due to the decreasing sampling
density of a magnitude-limited redshift survey at higher $z$, which
introduces a bias against identifying groups at higher redshift. Note
that our adopted stellar mass bins are well-matched to those of
\citet{iovino09}, taking into account differences in the adopted value
of the Hubble parameter.

To test whether our high-density subsample and the random low-density
subsamples are drawn from the same underlying stellar mass
distribution, we apply the one-sided Wilcoxon-Mann-Whitney (WMW) $U$
test \citep{mann47}, a non-parametric test that is highly robust to
non-Gaussianity because it relies on ranks rather than observed
values. The result of the WMW $U$ test is the probability ($P_U$) that
a value of the $U$ statistic equal to the observed value or more
extreme would result if the ``null'' hypothesis (that both samples are
drawn from identical distributions) holds. The WMW $U$ test is
particularly useful for small data sets (e.g., compared to other
related tests such as the chi-square two-sample test,
\citealt{wall03}), as we have when selecting galaxies from a narrow
stellar mass range and in extreme environments, due to its
insensitivity to outlying data points, its avoidance of binning, and
its high efficiency. Note that since this test is one-sided, possible
$P_{U}$ values range from $0$ to $0.5$; for a $P_{U}$ value below
$0.025$ (corresponding closely to $2\sigma$ for a Gaussian), we can
reject the null hypothesis (that the two samples have the same
distribution) at greater than $95\%$ significance.

\begin{figure*}[h]
\centering
\plotone{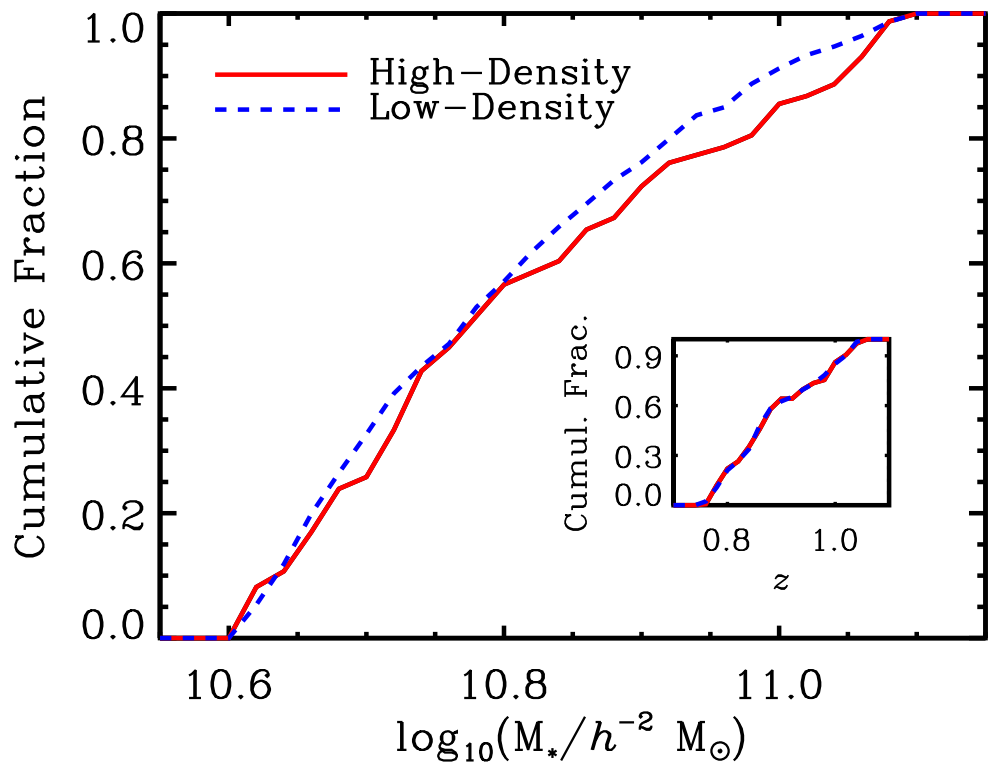}
\caption{the cumulative stellar mass and redshift (see inset)
  distributions for the $159$ DEEP2 galaxies comprising the top $10\%$
  of the environment distribution within the stellar mass and redshift
  ranges of $10.6 < \log_{10}({\rm M}_{*} / h^{-2}\ {\rm M}_{\sun}) <
  11.1$ and $0.75 < z < 1.05$ in comparison to the corresponding
  cumulative distributions for the $1000$ random galaxy subsamples
  drawn from the lowest $50\%$ of the same environment
  distribution. As discussed in the text, the low-density subsamples,
  which are each composed of $159$ galaxies, are selected to have the
  same redshift distribution as the high-density population. However,
  the stellar mass distribution is found to be biased in the
  different environment regimes.}
\label{mass_fig1}
\end{figure*}

In Figure \ref{mass_fig1}, we plot the cumulative distribution of
stellar masses for the $159$ sources in the high-density subsample
along side that for the 1000 random subsamples (each consisting of
$159$ galaxies) matched in redshift but residing in low-density
environments. Performing a one-sided WMW $U$ test on the stellar mass
measurements for the low- and high-density populations, we find that
the stellar mass distribution for the galaxies in high-density
environments is skewed to slightly higher stellar mass, with a
probability of $P_U \sim 0.09$. Meanwhile, the cumulative redshift
distributions for the low- and high-density subsamples, shown in the
inset of Figure \ref{mass_fig1}, are well-matched with the WMW $U$
test yielding a $P_U > 0.48$, thus confirming that the redshift
distributions for the two samples are indistinguishable and therefore
not significantly contributing to the difference in the stellar mass
distributions.

\begin{figure*}[tb!]
\centering
\plotone{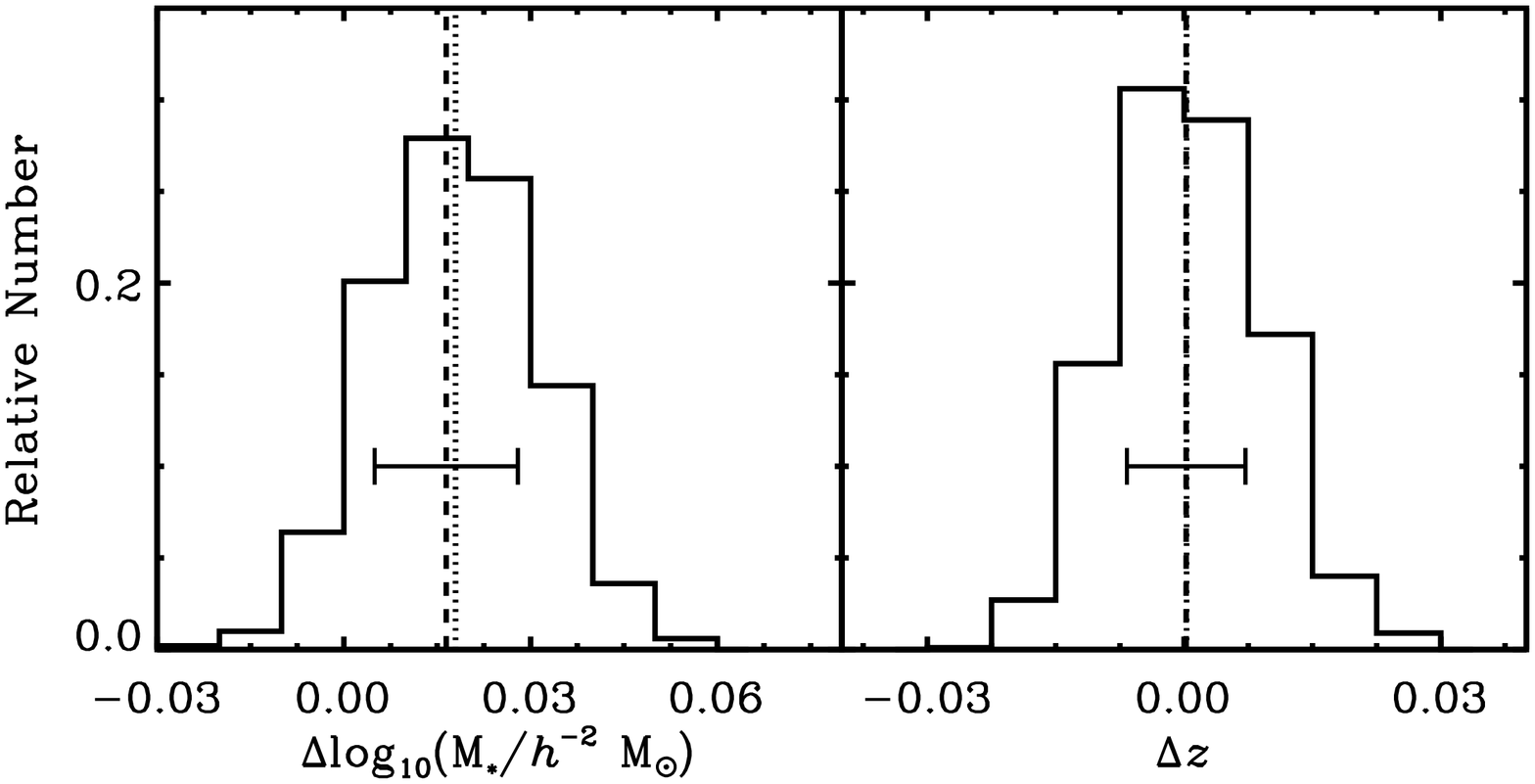}
\caption{the distribution of differences between the Hodges-Lehmann
  (H-L) estimator of the mean stellar mass (\emph{left}) and mean
  redshift (\emph{right}) for the high-density subsample relative to
  the corresponding H-L estimator of the mean for each of the $1000$
  low-density subsamples. The median differences in stellar mass and
  redshift are denoted by the dotted vertical lines, while the dashed
  vertical lines and corresponding error bars show the difference in
  the arithmetic means (and uncertainty in that difference) for the
  low-density and high-density populations. We find a small offset in
  stellar mass of $\Delta \log_{10}({\rm M}_{*} / h^{-2}\ {\rm
    M}_{\sun}) \sim 0.019$, while the difference in mean redshift of
  the two samples is consistent with zero (by construction). Within
  the broad stellar mass range of $10.6 < \log_{10}({\rm M}_{*} /
  h^{-2}\ {\rm M}_{\sun}) < 11.1$, we detect a weak stellar
  mass-environment relation at $z \sim 0.9$ such that more massive
  galaxies favor higher-density regions.}
\label{mass_fig2}
\end{figure*}

The results of the WMW $U$ test are supported by a comparison of the
mean stellar masses for the low- and high-density subsamples, which
are distinct at a $\gtrsim \! 1.5 \sigma$ level, thereby supporting
the conclusion that there is a non-negligible stellar mass-environment
relation within the given mass bin. In addition to directly comparing
the arithmetic means of the stellar mass distributions, we also
utilize the Hodges-Lehmann (H-L) estimator of the mean, which is given
by the median value of the mean stellar mass computed over all pairs
of galaxies in the sample \citep{hodges63}. Like taking the median of
a distribution, the H-L estimator of the mean is robust to outliers,
but, unlike the median, yields results with scatter (in the Gaussian
case) comparable to the arithmetic mean. Thus, by using the H-L
estimator of the mean, we gain robustness as in the case of the
median, but unlike the median, our measurement errors are increased by
only a few percent. In Figure \ref{mass_fig2}, we show the
distribution of differences between the Hodges-Lehmann estimator of
the mean stellar mass for the high-density subsample relative to that
for the $1000$ low-density subsamples, where the median difference in
stellar mass is $\Delta \log_{10}({\rm M}_{*} / h^{-2}\ {\rm
  M}_{\sun}) \sim 0.019$ as illustrated by the dotted vertical
line. Within the stellar mass range of $10.6 < \log_{10}({\rm M}_{*} /
h^{-2}\ {\rm M}_{\sun}) < 11.1$, we find (by all of these methods)
$\sim \! 1.5 \sigma$ evidence for a weak stellar mass-environment
relation at $z \sim 0.9$ such that more massive galaxies favor
higher-density regions.

This weak correlation between stellar mass and environment is robust
to variations in the stellar mass and redshift ranges used to select
the galaxy population. For example, pushing to slightly lower stellar
masses ($10.2 < \log_{10}({\rm M}_{*}/h^{-2}\ {\rm M}_{\sun}) <
10.7$), while still limiting the sample to a range of only $0.5$ dex,
we again find evidence for a weak correlation between between stellar
mass and environment (see Table \ref{res_tab1}). Expanding the range
of stellar masses probed to $10.1 < \log_{10}({\rm M}_{*}/h^{-2}\ {\rm
  M}_{\sun}) < 11.1$, the stellar mass-density relation is
increasingly evident, such that we find a highly-significant
correlation between stellar mass and environment at $z \sim 0.9$, with
$P_{U} < 0.01$. This result is in general agreement with existing
studies of the projected galaxy correlation function as a function of
stellar mass at $z \sim 1$ \citep[e.g.,][]{meneux08, foucaud10}, which
find weak evidence that more massive galaxies are more strongly
clustered on small scales. Likewise, measurements of the stellar mass
function at slightly lower redshift $(z \sim 0.5)$ also suggest a
variation in the shape of the mass function with local environment
\citep{bolzonella09, kovac09, peng10}.

Thus, when looking to quantify environmental dependencies at fixed
stellar mass, the use of broad stellar mass bins (e.g., those used by
\citealt{scodeggio09}, \citealt{iovino09}, \citealt{tasca09},
\citealt{kovac09}, and \citealt{grutzbauch10}) or simple mass-limited
sample selections \citep[e.g.,][]{vdw07, pannella09, rettura10} is
inappropriate as both approaches implicitly assume that there is no
variation in the stellar mass function with environment. In the
following subsection, we develop improved techniques that account for
the apparent environmental dependence of the shape of the stellar mass
function, thereby enabling an unbiased analysis of the color-density
relation at fixed stellar mass.

\subsection{The Color-Density Relation at Fixed Stellar Mass}
\label{sec_results2}

The ultimate goal of this work is not to study the correlation between
stellar mass and environment, but rather to investigate the
relationship between galaxy color and environment at fixed stellar
mass. But as shown in the above analysis, we must be careful to
account for the weak correlation between stellar mass and overdensity
even with stellar mass bins as narrow as $0.5$ dex. Thus, we now
select those galaxies within the top $10\%$ of the overdensity
distribution for all galaxies at $10.6 < \log_{10}({\rm M}_{*} /
h^{-2}\ {\rm M}_{\sun}) < 11.1$ and $0.75 < z < 1.05$ (the same
high-density subsample of $159$ galaxies), and from the corresponding
bottom $50\%$ of the overdensity distribution we randomly draw 1000
subsamples (each composed of $159$ galaxies) so as to match the joint
redshift and stellar mass distributions of the galaxies in the
high-density subsample. Members of the low-density subsample are drawn
randomly from within a two-dimensional window with dimensions of
$\Delta z^2 < 4 \cdot 10^{-4}$ and $\Delta \log_{10}({\rm M}_{*})^2 <
5 \cdot 10^{-5}$ of a randomly-selected object in the high-density
subsample. Varying the size of this window by factors of a few in each
dimension has no significant effect on our results. Given the random
nature of the matching, some objects are repeated in the low-density
subsamples. However, for each subsample of $159$ galaxies, $\sim \!
90\%$ of the galaxies are unique; requiring all members of a subsample
to be unique would skew the statistics \citep{efron81}. By matching
our high- and low-density subsamples in stellar mass as well as
redshift, we are able to effectively study the correlation between
galaxy properties such as color and environment at fixed stellar mass.

\begin{figure*}[tb!]
\centering
\plotone{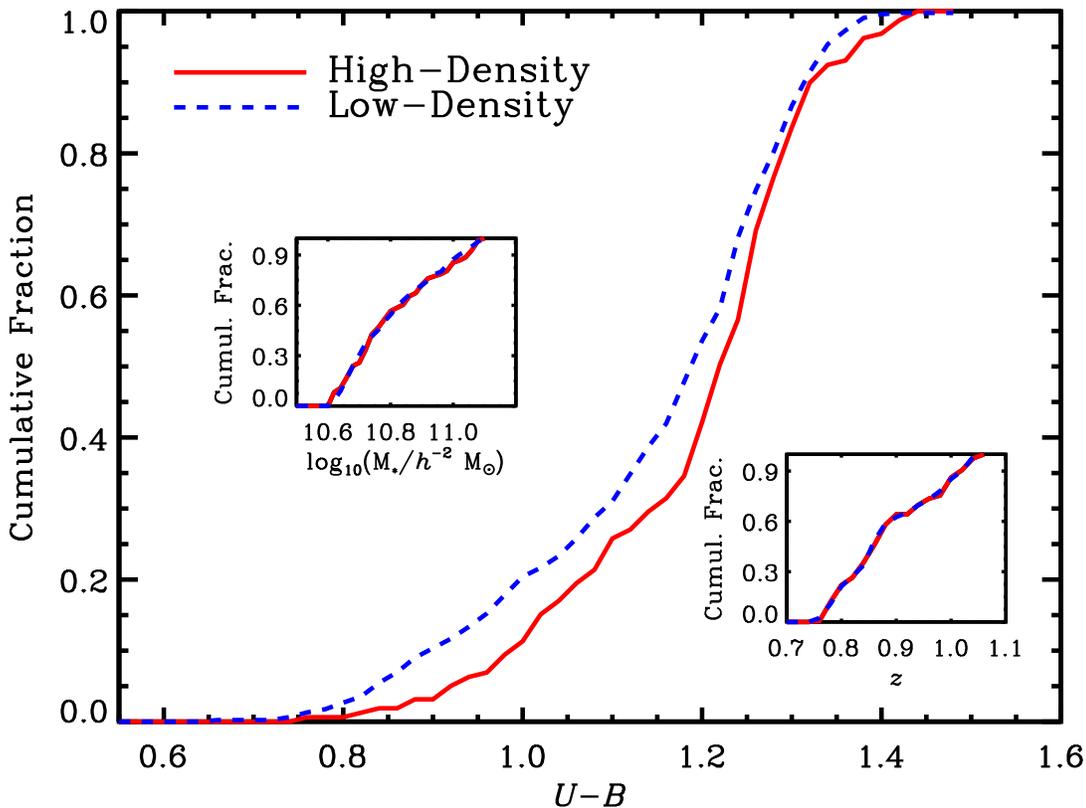}
\caption{the cumulative rest-frame $U-B$ color, stellar mass, and
  redshift distributions for the $159$ DEEP2 galaxies with the highest
  measured overdensities (the top $10\%$ of the environment
  distribution) within the stellar mass and redshift ranges of $10.6 <
  \log_{10}({\rm M}_{*} / h^{-2}\ {\rm M}_{\sun}) < 11.1$ and $0.75 <
  z < 1.05$ in comparison to the corresponding cumulative
  distributions for the $1000$ random galaxy subsamples drawn from the
  lowest $50\%$ of the same environment distribution. As discussed in
  the text, the low-density subsamples, which are each composed of
  $159$ galaxies, are selected to have the same stellar mass and
  redshift distributions as the high-density population (see inset
  plots). For galaxies with stellar masses of $10.6 < \log_{10}({\rm
    M}_{*} / h^{-2}\ {\rm M}_{\sun}) < 11.1$ at $z \sim 0.9$, we find
  a significant color-density relation at fixed stellar mass such that
  red galaxies preferentially reside in overdense regions relative to
  their bluer counterparts.}
\label{color_fig}
\end{figure*}

\begin{figure*}[tb!]
\centering
\plotone{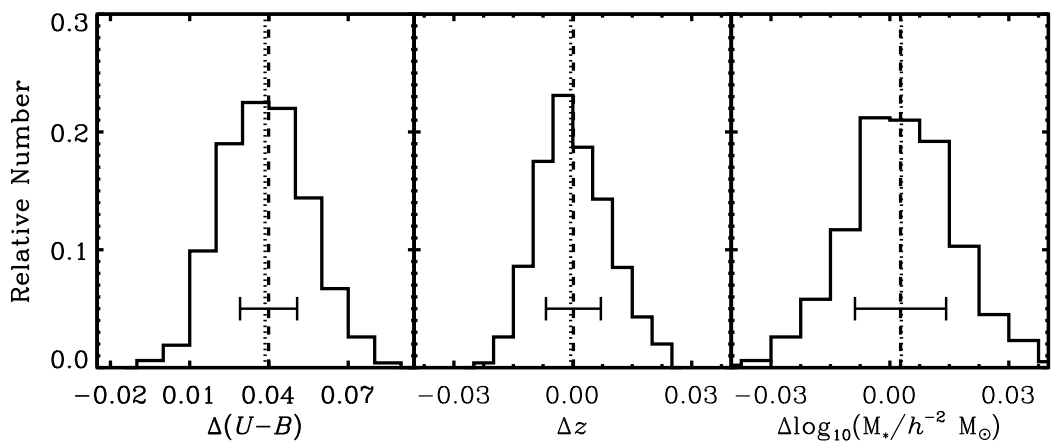}
\caption{the distribution of differences between the Hodges-Lehmann
  (H-L) estimator of the mean $U-B$ color, redshift, and stellar mass
  for the high-density subsample relative to the corresponding H-L
  estimator of the mean for each of the $1000$ low-density
  subsamples. The median differences in $U-B$, $z$, and stellar mass
  are denoted by the dotted vertical lines, while the dashed vertical
  lines and corresponding error bars show the difference in the
  arithmetic means (and uncertainty in that difference) for the
  low-density and high-density populations (taken as a whole). Recall
  that the redshift and stellar mass distributions of the low- and
  high-density samples match by construction. We find a significant
  offset in rest-frame color of $\Delta (U-B) \sim 0.04$, while the
  difference in mean redshift and stellar mass for the two samples is
  consistent with zero.}
\label{color_fig2}
\end{figure*}

As shown in Figures \ref{color_fig} and \ref{color_fig2}, we find a
significant relationship at $z \sim 0.9$ between rest-frame $U-B$
color and local galaxy overdensity at fixed stellar mass for galaxies
with $10.6 < \log_{10}({\rm M}_{*} / h^{-2}\ {\rm M}_{\sun}) <
11.1$. The cumulative color distributions for the two subsamples drawn
from separate environment regimes are significantly distinct, with the
galaxies in high-density environments skewed towards redder rest-frame
colors. The WMW $U$ test confirms what is apparent in Figure
\ref{color_fig}, yielding $P_{U} < 0.002$ when comparing the color
distributions of the high- and low-density subsamples: these color
distributions are distinct at $> \! 3\sigma$ confidence. By
construction, the corresponding stellar mass and redshift
distributions are indistinguishable (see inset plots in Fig.\
\ref{color_fig}), with $P_{U} \sim 0.47$ and $P_{U} \sim 0.49$,
respectively.

In Figure \ref{color_fig2}, we show the distribution of differences
between the Hodges-Lehmann estimator of the mean $U-B$ color, stellar
mass, and redshift for the high-density subsample relative to the
corresponding estimator of the mean for the $1000$ low-density
subsamples. The median differences in stellar mass and redshift are
consistent with zero, $\Delta z \sim -1 \cdot 10^{-4} \pm 0.009$ and
$\Delta \log_{10}({\rm M}_{*} / h^{-2}\ {\rm M}_{\sun}) \sim 0.003 \pm
0.012$, while the median offset in rest-frame color is $\Delta(U-B)
\sim 0.033 \pm 0.013$ such that galaxies in high-density environs are
typically redder in color at fixed stellar mass and
redshift. Similarly, the arithmetic means of the two color
distributions are accordingly found to be distinct at $\gtrsim \! 3.5
\sigma$ (see Table \ref{res_tab2}). Altogether, the DEEP2 data show a
robust color-density relation at fixed stellar mass at $z \sim 0.9$
for galaxies in the stellar mass regime of $10.6 < \log_{10}({\rm
  M}_{*} / h^{-2}\ {\rm M}_{\sun}) < 11.1$.

To test the robustness of our results to the particularities of the
sample selection, we repeat the analysis described above for several
samples spanning varying redshift and stellar mass regimes. For
example, broadening the redshift range over which we select galaxies
to $0.75 < z < 1.25$, thereby increasing the size of the sample, we
again find a statistically significant relationship between rest-frame
color and environment within our adopted stellar mass bin of $10.6 <
\log_{10}({\rm M}_{*} / h^{-2}\ {\rm M}_{\sun}) < 11.1$. For the $211$
galaxies in the high-density regime (again the highest $10\%$ of the
overdensity distribution) at $0.75 < z < 1.25$, the cumulative
distribution of $U-B$ color is skewed towards redder colors relative
to the comparison set of galaxies in low--density environments,
yielding $P_{U} < 0.01$ and with the means of the two color
distributions distinct at a $\sim \! 3\sigma$ level. In Table
\ref{res_tab2}, we list the results from similar analyses of several
other galaxy samples. When varying the redshift and/or stellar mass
regimes probed, we continue to find a significant color-density
relation at fixed stellar mass at $z \sim 1$. This result is found to
hold even at slightly lower masses; restricting to only those galaxies
with $10.2 < \log_{10}({\rm M}_{*} / h^{-2}\ {\rm M}_{\sun}) < 10.7$
(a sample with a median stellar mass of $\log_{10}({\rm M}_{*} /
h^{-2}\ {\rm M}_{\sun}) \sim 10.4$), we again find that redder
galaxies tend to favor overdense regions at fixed stellar mass and
redshift.

\begin{deluxetable*}{c c  c c c  c c c}
\tablewidth{0pt}
\tablecolumns{7}
\tablecaption{\label{res_tab1} Summary of Results for Stellar
  Mass-Density Relation at Fixed Redshift} 
\tablehead{ Sample & $N_{\rm high-density}$ & $P_{U}(z)$ & 
$P_{U}(\log_{10}({\rm M}_{*}))$ & 
$\Delta z$ & $\Delta \log_{10}({\rm M}_{*})$ }
\startdata
\hline
$0.75 < z < 1.05$ & \multirow{2}{*}{$159$} & 
\multirow{2}{*}{$0.49$} & \multirow{2}{*}{$0.09$} & 
\multirow{2}{*}{$0.000 \pm 0.007$} &
\multirow{2}{*}{$0.019 \pm 0.012$} \\
$10.6 < \log_{10}({\rm M}_{*}) < 11.1$ &  &  
  &  &  &  \\
\hline
$0.75 < z < 1.05$ & \multirow{2}{*}{$456$} & 
\multirow{2}{*}{$0.48$} & \multirow{2}{*}{$< \! 0.01$} & 
\multirow{2}{*}{$0.000 \pm 0.004$} &
\multirow{2}{*}{$0.051 \pm 0.012$} \\
$10.1 < \log_{10}({\rm M}_{*}) < 11.1$ &  &  
  &  &  &  \\
\hline
$0.75 < z < 1.25$ & \multirow{2}{*}{$211$} & 
\multirow{2}{*}{$0.49$} & \multirow{2}{*}{$0.13$} & 
\multirow{2}{*}{$0.000 \pm 0.005$} &
\multirow{2}{*}{$0.012 \pm 0.010$} \\
$10.6 < \log_{10}({\rm M}_{*}) < 11.1$ &  &  
  &  &  &  \\
\hline
$0.75 < z < 1.05$ & \multirow{2}{*}{$273$} & 
\multirow{2}{*}{$0.49$} & \multirow{2}{*}{$0.08$} & 
\multirow{2}{*}{$0.000 \pm 0.006$} &
\multirow{2}{*}{$0.012 \pm 0.008$} \\
$10.2 < \log_{10}({\rm M}_{*}) < 10.7$ &  &  
  &  &  &  \\
\hline
\enddata
\tablecomments{For a variety of galaxy samples, we list the results of
  the WMW $U$ test ($P_{U}$) and the difference in the arithmetic
  means computed from a comparison of the redshift and stellar mass
  distributions of the respective low- and high-density samples. The
  $P$-value, $P_{U}$, indicates the probability that differences in
  the distribution of the stated quantity as large as those observed
  (or larger) would occur by chance if the two samples shared
  identical distributions. The number of galaxies in the high-density
  sample (picked to be the top 10\% of the environment distribution)
  is given by $N_{\rm high-density}$. Note that the difference in the
  mean for the galaxy property $x$ is given by $<{x}_{\rm
    high-density}> - <{x}_{\rm low-density}>$ and that stellar masses
  are in units of $h^{-2}\ {\rm M}_{\sun}$.}
\end{deluxetable*}

\begin{deluxetable*}{c c  c c c  c c c}
\tablewidth{0pt}
\tablecolumns{7}
\tablecaption{\label{res_tab2} Summary of Results for Color-Density
  Relation at Fixed Stellar Mass and Redshift} 
\tablehead{ Sample & $N_{\rm high-density}$ & $P_{U}(z)$ & 
$P_{U}(\log_{10}({\rm M}_{*}))$ & 
$P_{U}(U-B)$ & $\Delta z$ & 
$\Delta \log_{10}({\rm M}_{*})$ & 
$\Delta(U-B)$}
\startdata
\hline
$0.75 < z < 1.05$ & \multirow{2}{*}{$159$} & 
\multirow{2}{*}{$0.49$} & \multirow{2}{*}{$0.47$} & 
\multirow{2}{*}{$< \! 0.01$} & \multirow{2}{*}{$0.000 \pm 0.007$} & 
\multirow{2}{*}{$0.002 \pm 0.012$} & \multirow{2}{*}{$0.040 \pm 0.011$} \\
$10.6 < \log_{10}({\rm M}_{*}) < 11.1$ &  &  
  &  &  &  & & \\
\hline
$0.75 < z < 0.95$ & \multirow{2}{*}{$101$} & 
\multirow{2}{*}{$0.48$} & \multirow{2}{*}{$0.47$} & 
\multirow{2}{*}{$0.01$} & \multirow{2}{*}{$0.000 \pm 0.005$} & 
\multirow{2}{*}{$0.002 \pm 0.015$} & \multirow{2}{*}{$0.043 \pm 0.014$} \\
$10.6 < \log_{10}({\rm M}_{*}) < 11.1$ &  &  
  &  &  &  & & \\
\hline
$0.75 < z < 1.25$ & \multirow{2}{*}{$211$} & 
\multirow{2}{*}{$0.47$} & \multirow{2}{*}{$0.46$} & 
\multirow{2}{*}{$< \! 0.01$} & \multirow{2}{*}{$0.000 \pm 0.008$} & 
\multirow{2}{*}{$0.002 \pm 0.010$} & \multirow{2}{*}{$0.032 \pm 0.010$} \\
$10.6 < \log_{10}({\rm M}_{*}) < 11.1$ &  &  
  &  &  &  & & \\
\hline
$0.75 < z < 1.05$ & \multirow{2}{*}{$456$} & 
\multirow{2}{*}{$0.48$} & \multirow{2}{*}{$0.46$} & 
\multirow{2}{*}{$< \! 0.01$} & \multirow{2}{*}{$0.000 \pm 0.004$} & 
\multirow{2}{*}{$0.003 \pm 0.012$} & \multirow{2}{*}{$0.029 \pm 0.009$} \\
$10.1 < \log_{10}({\rm M}_{*}) < 11.1$ &  &  
  &  &  &  & & \\
\hline
$0.85 < z < 1.15$ & \multirow{2}{*}{$404$} & 
\multirow{2}{*}{$0.47$} & \multirow{2}{*}{$0.49$} & 
\multirow{2}{*}{$0.02$} & \multirow{2}{*}{$0.000 \pm 0.004$} & 
\multirow{2}{*}{$0.002 \pm 0.013$} & \multirow{2}{*}{$0.019 \pm 0.010$} \\
$10.1 < \log_{10}({\rm M}_{*}) < 11.1$ &  &  
  &  &  &  & & \\
\hline
$0.75 < z < 1.05$ & \multirow{2}{*}{$273$} & 
\multirow{2}{*}{$0.49$} & \multirow{2}{*}{$0.47$} & 
\multirow{2}{*}{$< \! 0.01$} & \multirow{2}{*}{$0.000 \pm 0.006$} & 
\multirow{2}{*}{$0.000 \pm 0.008$} & \multirow{2}{*}{$0.032 \pm 0.011$} \\
$10.2 < \log_{10}({\rm M}_{*}) < 10.7$ &  &  
  &  &  &  & & \\
\hline
\enddata
\tablecomments{For a variety of galaxy samples, we list the results of
  the WMW $U$ test ($P_{U}$) and the difference in the arithmetic
  means computed from a comparison of the redshift, stellar mass, and
  color distributions of the respective low- and high-density
  samples. The $P$-value, $P_{U}$, indicates the probability that
  differences in the distribution of the stated quantity as large as
  those observed (or larger) would occur by chance if the two samples
  shared identical distributions. The number of galaxies in the
  high-density sample (picked to be the top 10\% of the environment
  distribution) is given by $N_{\rm high-density}$. Note that the
  difference in the mean for the galaxy property $x$ is given by
  $<{x}_{\rm high-density}> - <{x}_{\rm low-density}>$ and that
  stellar masses are in units of $h^{-2}\ {\rm M}_{\sun}$. Finally,
  note that the galaxy samples listed here are distinct from those
  detailed in Table \ref{res_tab1}. These samples are matched in
  redshift as well as stellar mass.}
\end{deluxetable*}

\section{Discussion}
\label{sec_disc}

As shown in Figure \ref{color_fig}, we find that there exists a
correlation between rest-frame $U-B$ color and local galaxy
overdensity within the high-mass ($\log_{10}({\rm M}_{*} / h^{-2}\
{\rm M}_{\sun}) \sim 10.8$) segment of the galaxy population at $z
\sim 0.9$ such that red galaxies favor higher-density environments at
fixed stellar mass. As noted in Section \ref{sec_intro}, several
recent studies utilizing data from VVDS and zCOSMOS have concluded
that no such correlation exists within this stellar mass and redshift
regime \citep[e.g.,][]{scodeggio09, iovino09}. Here, we discuss the
likely reasons for the discrepancy between these conclusions and our
results.

It is highly unlikely that the highly significant color-density
relation at fixed stellar mass within the DEEP2 data is
spurious. Rest-frame galaxy colors and stellar masses, which depend on
the adopted stellar population models, photometry, and only coarsely
on redshift, are entirely independent of the environment measures,
which depend upon angular position and high-precision redshift
information (both for the galaxy in question and neighboring
galaxies). There is no reasonable mechanisms that would produce a
false correlation between these independent quantities. For instance,
one might suggest that contamination of photometry is an issue in
dense regions; however, even in the most overdense environments (e.g.,
the top $5\%$ of the environment distribution), the typical distance
to the $3^{\rm rd}$-nearest neighbor corresponds to $\sim \!
35^{\prime\prime}$ on the sky, which is much larger than the aperture
sizes used in photometry. Instead, as highlighted in Section
\ref{sec_intro}, it is far more likely that the color-density relation
apparent in DEEP2 has been smeared out in studies using other data
sets due to the smaller sample sizes employed and the significantly
larger errors in the environment measures derived from those data
sets.

Many of the limitations regarding the VVDS data set, as they relate to
detecting correlations between galaxy properties and environment at $z
\sim 1$, are discussed at length by \citet{cooper07} in a comparison
to the work by \citet{cucciati06}. Using DEEP2 data, \citet{cooper07}
find a significant correlation between rest-frame galaxy color and
environment at $M_B < -20.5 - 5 \cdot \log_{10}(h)$ and $0.9 < z <
1.2$, where none is evident in the analysis of VVDS data presented by
\citet{cucciati06}. While both of these studies employed
luminosity-selected galaxy samples, the lessons learned from the
comparison are no less applicable to analyses of samples selected
according to stellar mass. Here, we review the arguments presented by
\citet{cooper07} as they relate to the more recent analyses using
VVDS, zCOSMOS, and other data sets.

At $z \gtrsim 0.75$, the sample size of both VVDS and zCOSMOS are
significantly smaller than that collected by DEEP2. At these
redshifts, DEEP2 has more than $26,891$ secure ($95\%$ or greater
confidence) redshifts, while the VVDS and zCOSMOS data sets include
only $1,581$ and $967$ high-quality (i.e., $\rm{flag} \! = \!  3,4$)
redshifts, respectively.

In addition to their smaller statistical power relative to DEEP2, both
VVDS and zCOSMOS have a significantly lower number density of tracer
objects \citep[cf.][]{newman09}, which directly influences the level
to which the resulting data sets can trace local galaxy
density. Crudely, errors on overdensity measurements on some length
scale should be dominated by Poisson statistics and hence should be
proportional to the square root of the number density of tracers. To
test the impact of a reduction in the sampling density, we dilute the
DEEP2 data set by a factor of $3$, thereby creating a sample that
mimics the sampling density of VVDS and zCOSMOS at $z \sim 1$. Using
this tracer population, we then recompute the local overdensity about
each galaxy in the full DEEP2 sample. Note that this approach allows
us to test the role of sampling density independent of sample
size. Performing the same analysis described in \S \ref{sec_results2},
but using environment measures computed with the diluted tracer
population, we still find a color-density relation at fixed stellar
mass. However, the significance of the trend is noticeably reduced.
For galaxies with $10.6 < \log_{10}({\rm M}_{*} / h^{-2}\ {\rm
  M}_{\sun}) < 11.1$ and $0.75 < z < 1.05$, we find that the color
distributions for galaxies in high- and low-density environments are
distinct with a $P_{U} \sim 0.03$ when using the diluted tracer
population versus $P_{U} < 0.01$ when measuring environment with the
full DEEP2 sample. Comparing the mean rest-frame colors as a function
of environment, we find that the difference in the arithmetic means,
$\Delta(U-B)$, between high- and low-density regions is significant at
only $\sim \! 2\sigma$ when employing the diluted tracer sample
compared to $\sim \! 4\sigma$ when the full data set is used to define
environment. Thus, while variation in the sampling density may not
entirely explain the lack of evidence for a correlation between
environment and rest-frame galaxy color at fixed stellar mass in the
VVDS and zCOSMOS data sets, it clearly plays a substantial role. As
the number density of tracers decreases, the uncertainty on each
environment measure increases such that underlying correlations with
environment are increasingly smeared out.

Analyses using the VVDS and zCOSMOS data sets also often rely on the
use of less accurate redshifts. For example, the VVDS redshift catalog
is dominated by lower-quality redshifts \citep[$\rm{flag} \! = \!
2$,][]{ilbert05} at $z \sim 1$. A number of tests have found that the
${\rm flag}=2$ redshifts have a non-negligible error rate \citep[$\sim
\! 20\%$, C.\ Wolf, private communication;][]{lefevre05}. The
inclusion of these sources in the analysis of \citet{scodeggio09}
results in both inaccurate rest-frame colors (via incorrect
$K$-corrections) as well as inaccurate overdensity measurements in
their high-redshift $(0.7 < z < 1.4)$ sample and thus contributes to
the lack of color-density relation observed within the $10.2 <
\log_{10}({\rm M}_{*}/h^{-2}\ {\rm M}_{\sun}) < 10.7$ stellar mass
regime \citep[see Fig.\ 5 of][]{scodeggio09}. Along the same lines,
\citet{tasca09} utilize zCOSMOS redshifts with quality codes ${\rm
  flag} \! = \! 4$, $3$, $9$, and $2$. This includes objects for which
only a single emission line was identified in the observed
spectrum. While these spectroscopic redshifts are checked with
photometric redshifts and comprise only $\sim \! 6\%$ of the entire
sample, these objects play a significantly larger role at $z > 0.8$,
comprising nearly one quarter of the sample at high redshift.

Given that there are multiple ways to explain the smearing out of a
real correlation between galaxy color and environment, but no
reasonable mechanism by which to generate a false correlation, we
conclude that the color-density relation at fixed stellar mass that we
observe is genuine. The authors of several recent papers appear to
have gone too far in concluding that because they detect no relation,
none exists; this has led to a problematic interpretation of their
results.

\section{Summary}
\label{sec_summary}

Herein, we use data from the DEEP2 Galaxy Redshift Survey to complete
a detailed study of the relationship between rest-frame galaxy color
and local environment at fixed stellar mass at intermediate
redshift. Our principal result is that at fixed stellar mass and
redshift we find a significant relationship between rest-frame $U-B$
color and local galaxy density at $z \sim 0.9$ within the massive
($10.6 < \log_{10}({\rm M}_{*}/h^{-2}\ {\rm M}_{\sun}) < 11.1$) galaxy
population. This color-density relation is such that red galaxies at a
given stellar mass are preferentially found in overdense regions,
thereby showing that the general pattern of environmental dependence
of galaxy properties at fixed stellar mass in the local Universe
\citep[e.g.,][]{kauffmann04, baldry06, vdw08} persists to $z \sim
1$. Our findings disagree with some recent results, which have found
no correlation between galaxy properties and environment at fixed
stellar mass at $z \gtrsim 0.8$ \citep[e.g.,][]{tasca09, iovino09,
  scodeggio09}.

As shown in this work, being unable to detect any correlation between
galaxy properties and environment within a specific data set does not
alone indicate a true lack of environmental dependence. Uncertainties
and limitations associated with observations are generally such that
they smear out the underlying correlation between environment and
galaxy properties; as a result, the observed strength of relationships
such as the color-density relation are lower limits to the inherent
environmental dependence. In order to glean anything meaningful from
an observed lack of correlation between environment and galaxy
properties such as color and morphology, additional tests must be
undertaken to illustrate that the lack of correlation is inconsistent
with being due to observational uncertainties
\citep[e.g.,][]{cooper07, gerke07}. Moreover, even an observed decline
or weakening in the strength of an environmental dependency with
redshift must be treated carefully; the observed amplitude of
environment trends will weaken simply because errors generally
increase and sample sizes generally decrease with increasing
redshift. Much of the recent environment- or clustering-related work
at $z \sim 1$ \citep[e.g.,][]{cucciati06, lefevre07, tasca09,
  iovino09, scodeggio09, kovac09, bolzonella09} omits such analysis,
making the associated conclusions difficult to interpret.

Ultimately, in this work, we have shown that the evolutionary state of
a galaxy at $z \sim 1$ as traced by rest-frame $U-B$ color (i.e.,
specific star-formation rate) is not solely determined by the stellar
mass of the galaxy. That is, our results indicate that environment
plays a role (or at the least is correlated with a causal physical
parameter such as dark matter halo mass) in defining the evolutionary
history of a galaxy at $z > 1$. As previously stated, existing
analyses in the local Universe arrive at a similar conclusion with
regard to the separable roles of stellar mass and environment at $z >
0$. We now know that the typically redder colors of galaxies in
high-density regions at $z < 1$ is not solely attributable to the
tendency of these extreme environs to host systematically more massive
galaxies. Instead, the variation in color with environment appears to
also be partially the result of density-dependent
effects. Establishing the relative importance of these two
evolutionary drivers (stellar mass and environment) is a difficult
task, requiring a careful examination of the associated measurement
errors. Such analysis will no doubt be the focus of much future work.

\vspace*{0.25in} 
%%%%%%%%%%%%%%%%%%%%%%%
%%% Acknowledgments %%%
%%%%%%%%%%%%%%%%%%%%%%%

\acknowledgments Support for this work was provided by NASA through
the Spitzer Space Telescope Fellowship Program. This work was also
supported in part by NSF grants AST-0507428, AST-0507483, AST-0071048,
AST-0071198, AST-0808133, and AST-0806732 as well as {\it Hubble Space
  Telescope} Archival grant, HST-AR-10947.01. KB acknowledges support
for this work provided by NASA through Hubble Fellowship grant
\#HF-01215, awarded by the Space Telescope Science Institute, which is
operated by the Association of Universities for Research in Astronomy,
Inc., for NASA, under contract NAS 5-26555. MCC thanks Greg Wirth and
the entire Keck Observatory staff for their help in the acquisition of
the DEEP2 Keck/DEIMOS data.

We also wish to recognize and acknowledge the highly significant
cultural role and reverence that the summit of Mauna Kea has always
had within the indigenous Hawaiian community. It is a privilege to be
given the opportunity to conduct observations from this mountain.

{\it Facilities:} \facility{Keck:II (DEIMOS)}

%%%%%%%%%%%%%%%%%%%%
%%% Bibliography %%%
%%%%%%%%%%%%%%%%%%%%
%\bibliographystyle{apj}
%\bibliography{apj-jour,letter}

\end{document}